\documentclass[twocolumn,showpacs]{revtex4}
\usepackage{epstopdf}
\usepackage{graphicx}
\usepackage{color}
\usepackage{epsf}
\usepackage{dcolumn}
\usepackage{bm}

\newcommand{\pd}[2]{\frac{\partial #1}{\partial #2}}

\newcommand{\omax}{\widetilde{\omega}_x}
\newcommand{\omay}{\widetilde{\omega}_y}

\newcommand{\rvec}{\bf r}
\newcommand{\vel} {\bf v}

\newcommand{\Ovec}{\bf \Omega}
\newcommand{\phase}{\phi}
\newcommand{\drho}{\delta \rho}
\newcommand{\dphase}{\delta \phase}
\newcommand{\vectwo}[2]{\left[ \begin{array}{c} #1 \\ #2 \end{array}\right]}
\newcommand{\matfour}[4]{\left[ \begin{array}{cc} #1 & #2 \\ #3 & #4
\end{array}\right]}

\newcommand{\ltsimeq}{\raisebox{-0.6ex}{$\,\stackrel
        {\raisebox{-.2ex}{$\textstyle <$}}{\sim}\,$}}
\newcommand{\gtsimeq}{\raisebox{-0.6ex}{$\,\stackrel
        {\raisebox{-.2ex}{$\textstyle >$}}{\sim}\,$}}

\def\gappeq{\mathrel{ \rlap{\raise.5ex\hbox{$>$}}
                      {\lower.5ex\hbox{$\sim$}}  } }
\def\lappeq{\mathrel{ \rlap{\raise.5ex\hbox{$<$}}
                      {\lower.5ex\hbox{$\sim$}}  } }

\begin{document}

\preprint{PRA}

\title{Instabilities leading to vortex lattice formation in rotating Bose-Einstein condensates}

\author{N. G. Parker$^{1}$, R. M. W. van Bijnen$^{1,2}$, and A. M.  Martin$^1$}

\address{$^{1}$ School of Physics, University of Melbourne, Parkville, Victoria
3010, Australia}
\address{$^{2}$ Eindhoven University of Technology, PO Box 513, 5600 MB Eindhoven, The Netherlands}

\date{\today}

\begin{abstract}
We present a comprehensive theoretical study of vortex lattice
formation in atomic Bose-Einstein condensates confined by a rotating
elliptical trap.  We consider rotating solutions of the classical
hydrodynamic equations, their response to perturbations, as well as
time-dependent simulations.  We discriminate three distinct,
experimentally testable, regimes of instability: {\em ripple}, {\em
interbranch}, and {\em catastrophic}. Under symmetry-breaking
perturbations these instabilities lead to lattice formation even at
zero temperature.  While our results are consistent with previous
theoretical and experimental results, they shed new light on lattice
formation.
\end{abstract}

 \maketitle

Vortex lattices are a striking manifestation of superfluidity in
rotating systems.  In recent years such states have been generated
in atomic Bose-Einstein condensates (BECs) via rotation in an
elliptical harmonic trap \cite{dalibard1,dalibard2,hodby}, in
analogy to the classic rotating bucket experiment in superfluid
Helium \cite{osborn}. Vortex lattices form for trap rotation
frequencies $\Omega$ in the region of $\Omega \sim 0.7
\omega_\perp$, where $\omega_\perp$ is the mean harmonic trap
frequency in the rotating plane.  Although a vortex lattice is
thermodynamically favourable for much lower rotation frequencies,
the {\em instabilities} necessary to seed vortex lattice formation
are only present in the region of $\Omega\sim 0.7\omega_\perp$. Such
instabilities have been predicted by hydrodynamic studies of
condensate solutions in the rotating frame \cite{Recati} and the
dynamical perturbations of these states \cite{Sinha}, and have also
been analysed in dipolar BECs \cite{dipolar}. Although numerical
simulations of the Gross-Pitaevskii equation have also observed
vortex lattice formation in this region
\cite{lobo,Emil,Nick1,Nick2,penckwitt,tsubota}, the results are
mixed: for $\Omega \ltsimeq 0.7\omega_\perp$, Refs.
\cite{penckwitt,lobo,tsubota} require thermal effects to enable
vortex nucleation, while Ref. \cite{Emil} does not; for $\Omega
\gtsimeq 0.7 \omega_\perp$, Refs. \cite{Emil,lobo,Nick1,Nick2}
observe a shape instability before nucleating vortices even in the
absence of thermal effects. This is consistent with experiments
\cite{hodby,abo} which indicate that lattice formation is
temperature-independent. Breaking the rotational symmetry of such
simulations has been shown to be crucial to enable realistic
nucleation of vortices \cite{Nick1}.

In this paper we present a thorough investigation of the
instabilities leading to vortex lattice formation in elliptical
traps in the regime $\Omega \leq\omega_\perp$.  We primarily
consider the cases when the trap ellipticity or rotation frequency
is introduced adiabatically, for which it is appropriate to
transform to the rotating frame. In order to probe the condensate
instabilities we consider the rotating frame condensate solutions
within the classical hydrodynamic regime and the response of these
states to dynamical perturbations. Furthermore we perform
time-dependent (TD) condensate simulations to test our results, and
probe the fate of the unstable condensate.

In the experiments \cite{dalibard1,hodby} the harmonic trap
confining the BEC is transformed radially into an ellipse and
rotated about the $z$-axis. In the limit of zero temperature, the
condensate can be approximated by a mean-field `wavefunction'
$\psi$, which can be expressed as $\psi = \sqrt{\rho} e^{i \phase}$,
where $\rho$ and $\phi$ are the condensate density and phase.  In
the frame rotating at $\Omega$, the density $\rho$ and fluid
velocity ${\bf v}=(\hbar/m)\nabla \phi$ \cite{speed} satisfy the
hydrodynamic equations,
\begin{equation}\label{hydrodynamic1}
\pd{\rho}{t} + \nabla \cdot \left[\rho(\vel-\Ovec \times \rvec)
\right] = 0,
\end{equation}
\begin{eqnarray}
m \pd{\vel}{t}+\nabla \left(\frac{1}{2} m {\bf v} \cdot {\bf
v}+\frac{1}{2}m(\omega_x^2x^2+\omega_y^2y^2+\omega_z^2z^2)\right.
\nonumber
\\
\left.+\rho g-\frac{\hbar^2}{2m}\frac{\nabla^2
\sqrt{\rho}}{\sqrt{\rho}}- m{\bf v} \cdot \left[\Ovec \times
\rvec\right] \right)=0. \label{hydrodynamic2}
\end{eqnarray}
Here $m$ is the atomic mass and $g=4\pi \hbar^2a/m$ is the
interaction coefficient, where $a$ is the $s$-wave scattering length
(in this work $g>0$). The harmonic trap is defined by the
frequencies $\omega_x$, $\omega_y$ and $\omega_z$. When the trap is
elliptical in the radial plane we can write the $x, y$ frequencies
as $\omega_x=\sqrt{1-\epsilon}~\omega_\perp$ and
$\omega_y=\sqrt{1+\epsilon}~\omega_\perp$, where $\epsilon$
characterises the trap ellipticity, and
$\omega_\perp^2=(\omega_x^2+\omega_y^2)/2$.

Following the approach by Recati {\it et al.} \cite{Recati}, we
employ the Thomas-Fermi (TF) approximation by neglecting the
$(\nabla^2 \sqrt{\rho}/\sqrt{\rho})$ term in
Eq.~(\ref{hydrodynamic2}). Furthermore we assume an irrotational
quadrupolar velocity field \cite{dum},
\begin{equation}\label{velocity}
{\bf v}= \alpha \nabla (xy),
\end{equation}
where $\alpha$ is the velocity field amplitude. Substituting this
into Eq.~(\ref{hydrodynamic2}) and setting $\partial \rho/\partial
t=\partial {\bf v}/\partial t=0$, we obtain stationary density
solutions of the form,
\begin{equation}
\rho_0= \frac{1}{g} \left[\mu - \frac{1}{2}m\left(\omax^2 x^2 +
\omay^2 y^2 + \omega_z^2 z^2 \right)\right],
\label{tfdensity}
\end{equation}
in the region where the chemical potential $\mu\geq m(\omax^2 x^2 +
\omay^2 y^2 + \omega_z^2 z^2)/2$, and $\rho_0=0$ elsewhere. The
effect of the rotation is to give {\em effective} trap frequencies,
which satisfy,
\begin{equation}
\omax^2= \left[(1 - \epsilon)+ \alpha^2 - 2 \alpha
\Omega\right]\omega_\perp^2 \label{omax}
\end{equation}
\begin{equation}
\omay^2= \left[(1 + \epsilon)+ \alpha^2 + 2 \alpha
\Omega\right]\omega_\perp^2, \label{omay}
\end{equation}
where $\alpha$ determines the ellipticity of the BEC density
\cite{Recati}. Plugging Eq.~(\ref{tfdensity}) into
Eq.~(\ref{hydrodynamic1}) we obtain \cite{Recati},
\begin{equation}\label{alphacontinuity}
\omax^2 (\alpha + \Omega) + \omay^2 (\alpha - \Omega) = 0,
\end{equation}
which specifies the stationary condensate solutions in the frame
rotating at $\Omega$ (in the TF regime).

\begin{figure}
\includegraphics[width=7.5cm]{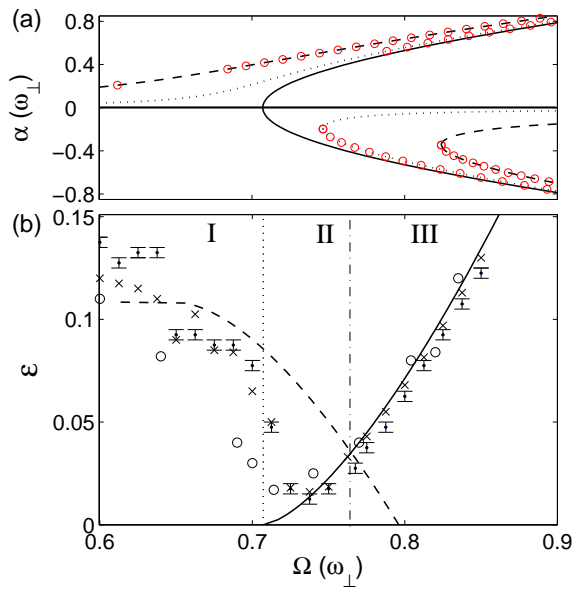}
\caption{(a) Velocity field amplitude $\alpha$ of the stationary
solutions of Eq.~(\ref{alphacontinuity}) as a function of rotation
frequency $\Omega$ for $\epsilon=0$ (solid line), $0.02$ (dotted
line) and $0.1$ (dashed line).  Regions of dynamical instability for
$\epsilon=0.02$ and $0.1$ are shown (circles). (b) Phase diagram of
$\epsilon$ versus $\Omega$.  Plotted are the bifurcation point
$\Omega_{\rm bif}(\epsilon)$ (solid line) from
Eq.~(\ref{alphacontinuity}), the onset of dynamical instability
$\Omega_{\rm ins}(\epsilon)$ from Eq.~(\ref{pertoperator})(dashed
line), and experimental data of Hodby {\it et al.} \cite{hodby}
(circles). Furthermore, time-dependent simulations of
Eqs.~(\ref{hydrodynamic1}) and (\ref{hydrodynamic2}) show the
critical ellipticities beyond which the condensate deviates from an
elliptical shape $\epsilon_{\rm cr}^{\rm dev}$ (crosses) and beyond
which lattice formation ultimately occurs $\epsilon_{\rm cr}^{\rm
inst}$ (points with error bars). The bifurcation point for
$\epsilon=0$ (dotted line) and crossing frequency $\Omega_{\rm X}$
(dot-dashed line) at which $\Omega_{\rm bif}(\epsilon)=\Omega_{\rm
ins}(\epsilon)$ are indicated. We assume a condensate with
$\mu=10\hbar \omega_\perp$. } \label{fig1}
\end{figure}

Figure 1(a) shows the stationary solutions in $\Omega$-$\alpha$
space for various values of trap ellipticity $\epsilon$.  In the
limit of $\epsilon=0$ (solid line) a non-rotating ($\alpha=0$)
solution occurs for all $\Omega$, with two additional solutions
bifurcating from the $\alpha=0$ axis at $\Omega^0_{\rm
bif}=\omega_\perp/\sqrt{2}$. For finite $\epsilon$ (dotted and
dashed lines), the $\alpha=0$ solution disappears and the plot
consists of two distinct branches. The {\em upper} branch (positive
$\alpha$) is single-valued and exists for all $\Omega$, while the
{\em lower} branch (negative $\alpha$) is double-valued and exists
only when $\Omega$ is greater than the bifurcation frequency
$\Omega_{\rm bif}(\epsilon)$.  As $\epsilon$ is increased from zero,
the branches move away from the $\alpha=0$ axis, as can be seen in
Fig.~\ref{fig1}(a). Furthermore the bifurcation point of the lower
branch $\Omega_{\rm bif}(\epsilon)$ shifts to higher rotation
frequencies, as shown in Fig.~\ref{fig1}(b)(solid line). The branch
diagrams, which have been probed experimentally \cite{dalibard2},
are key to understanding the response of the system to the adiabatic
introduction of trap ellipticity $\epsilon$ or rotation frequency
$\Omega$. Before any rotation/ellipticity is applied the BEC has
$\alpha=0$. When $\Omega$ is increased adiabatically (for fixed
$\epsilon$), the BEC follows the upper branch, with increasing
$\alpha$ and ellipticity in the density profile. When $\epsilon$ is
introduced adiabatically (for fixed $\Omega$), the BEC can follow
two routes, depending on the value of $\Omega$ relative to the
bifurcation point $\Omega^0_{\rm bif}$ (dotted line in
Fig.~\ref{fig1}(b)). For $\Omega<\Omega^0_{\rm bif}$, the lower
branch is nonexistent and the BEC follows the upper branch to
increasing values of $\alpha$. For $\Omega>\Omega^0_{\rm bif}$, the
lower branch moves from $\alpha=0$ to negative $\alpha$, and the BEC
follows this route. However, as $\epsilon$ is increased, the edge of
the lower branch $\Omega_{\rm bif}(\epsilon)$ shifts to higher
frequencies, and when $\Omega_{\rm bif}(\epsilon)>\Omega$ the lower
branch no longer exists. In this manner, the evolution of the
branches can induce instability, and has been linked to lattice
formation \cite{dalibard2}.

The stationary solutions of Eq.~(\ref{alphacontinuity}) are not
necessarily {\em stable} solutions. To investigate their stability
we follow the approach of Sinha and Castin \cite{Sinha} by
considering small perturbations $\delta \rho$ and $\delta \phi$ of
stationary solutions of density $\rho_0$ and phase $\phi_0$. Taking
variational derivatives of Eqs.~(\ref{hydrodynamic1}) and
(\ref{hydrodynamic2}) we obtain the time evolution equations,
\begin{equation}\label{pertoperator}
\pd{}{t}\vectwo{\dphase}{\drho} = -\matfour{\vel_c \cdot
\nabla}{g/m}{\nabla \cdot (\rho_0 \nabla)}{\ \vel_c \cdot \nabla}
\vectwo{\dphase}{\drho}
\end{equation}
where $\vel_c = \vel - \Ovec \times \rvec$ is the velocity field in
the rotating frame. The eigenfunctions of Eq.(\ref{pertoperator})
grow in time as ${\rm e}^{\lambda t}$, where $\lambda$ is the
corresponding eigenvalue. Solutions are unstable when there exist
one or more eigenfunctions for which $\lambda$ contains a positive
real part. Such unstable solutions are thought to seed vortex
lattice formation \cite{Sinha}.

Using a polynomial ansatz for the perturbations, this method
predicts a region of dynamical instability when $\Omega$ exceeds a
critical value $\Omega_{\rm ins}(\epsilon)$ \cite{Sinha}, as
indicated in Fig.~\ref{fig1}(a) (circles). The unstable modes have
lengthscales of the order of the condensate size, much greater than
the healing length $\xi=\hbar/\sqrt{mn_0 g}$ which characterises the
vortex size. In the limit of $\epsilon=0$, $\Omega_{\rm ins}\approx
0.78 \omega_\perp$. As $\epsilon$ is increased, $\Omega_{\rm
ins}(\epsilon)$ is reduced (dashed line in Fig.~\ref{fig1}(b)). Note
that the outlying point in Fig.~\ref{fig1}(a) for $\epsilon=0.1$
(dashed line) at $\Omega\approx 0.61\omega_\perp$ is not considered
to be in the region of instability due to its narrow width and
comparatively small eigenvalues \cite{narrow}. Note that, for the
parameter space of interest, on the lower branch the solution
closest to the $\alpha=0$ axis is never dynamically unstable. A key
rotation frequency in our work is $\Omega_{\rm X}$, which is the
crossing point of $\Omega_{\rm bif}(\epsilon)$ and $\Omega_{\rm
ins}(\epsilon)$, and has the value $\Omega_{\rm X}\approx
0.765\omega_\perp$ (dot-dashed line in Fig.~\ref{fig1}(b)).

Based on the stationary solutions of Eq.~(\ref{alphacontinuity}) and
the dynamical instability of Eq.~(\ref{pertoperator}) we can predict
the stability of a BEC following the adiabatic introduction of
$\Omega$ or $\epsilon$. However, to determine how the instability
{\em manifests} itself and whether it leads to lattice formation we
perform TD simulations of Eqs.~(\ref{hydrodynamic1}) and
(\ref{hydrodynamic2}) in the laboratory frame. This is equivalent to
solving the Gross-Pitaevskii equation \cite{Nick1}. Following
previous approaches \cite{Nick1,Emil,tsubota,penckwitt}, and noting
that the solutions of Eqs.~(\ref{alphacontinuity}) and
(\ref{pertoperator}) are independent of $z$ \cite{3d}, we consider a
2D system. The initial state is found by imaginary-time propagation.
Then, in real time, either $\epsilon$ is ramped up linearly at a
rate of $d\epsilon/dt=10^{-4}\omega_\perp$ (for fixed $\Omega$), or
$\Omega$ is ramped up linearly at a rate
$d\Omega/dt=10^{-2}\omega_\perp^2$ (for fixed $\epsilon$). We
monitor the evolution of $\alpha$ by fitting the velocity field in a
central region $[3\times3]\mu m$ to the form of
Eq.~(\ref{velocity}).

In such `idealised' simulations the trap and BEC can maintain a
two-fold rotational symmetry to unrealistic levels. This strongly
inhibits vortex nucleation since the vortices must enter in pairs
rather than individually.  Symmetry-breaking, rather than thermal
effects, has been shown to the crucial in simulating lattice
formation \cite{Nick1}. Indeed, Eq.~(\ref{pertoperator}) predicts
that only odd modes of the system are dynamically unstable. Previous
studies \cite{penckwitt,lobo} have broken this symmetry through the
introduction of thermal `noise'. To overcome this problem we shift
the trap by half a healing length before running in real-time,
thereby allowing excitation of odd modes \cite{numerical_noise}.

We first consider the adiabatic increase of trap ellipticity
$\epsilon$ for fixed $\Omega$.  We discriminate three cases of
instability, which each occur in distinct frequency regimes, as
indicated in Fig.~\ref{fig1}(b): (I) ripple instability, (II)
interbranch instability, and (III) catastrophic instability. We will
discuss each in turn.

\begin{figure}
\centering \vspace{0.3cm}
\includegraphics[width=3.32cm,clip=true]{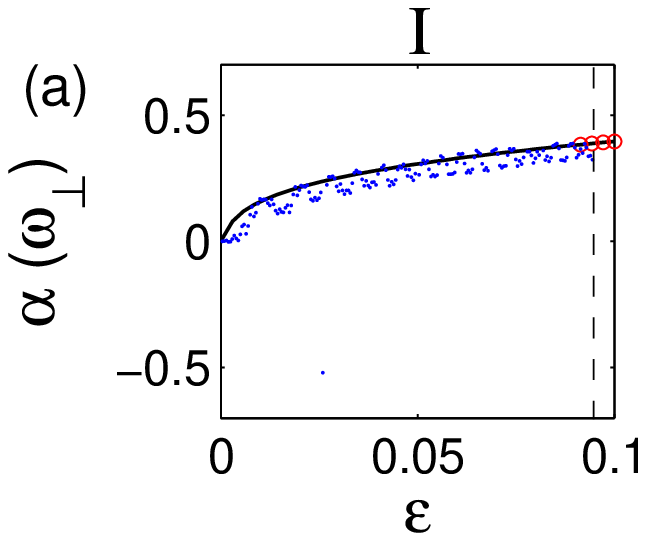}
\hspace{-0.15cm}
\includegraphics[width=2.6cm,clip=true]{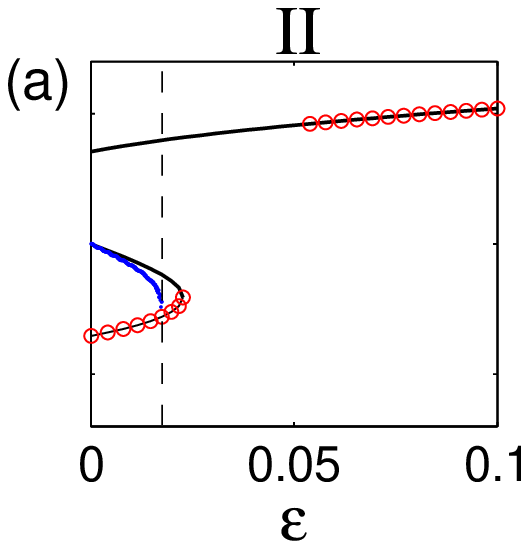}
\hspace{-0.15cm}
\includegraphics[width=2.63cm,clip=true]{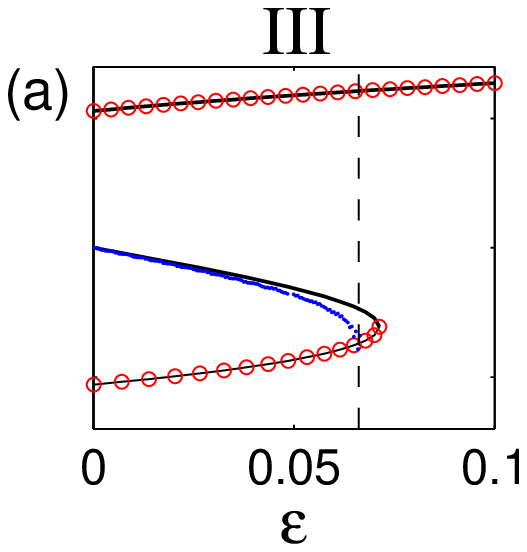}
\\
\hspace{0.4cm}
\includegraphics[width=2.2cm,clip=true]{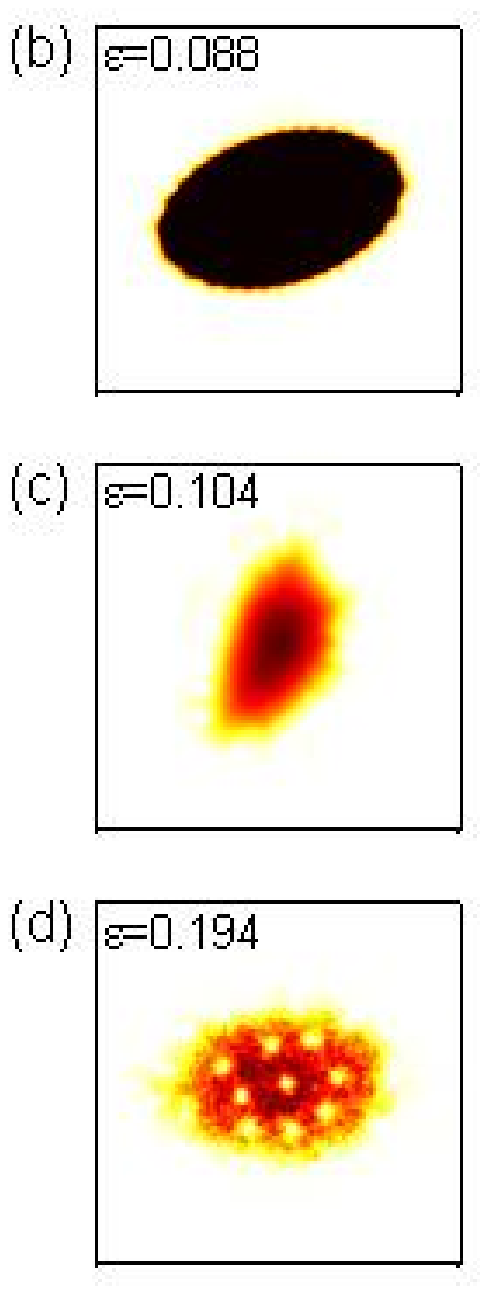}
\hspace{0.2cm}
\includegraphics[width=2.2cm,clip=true]{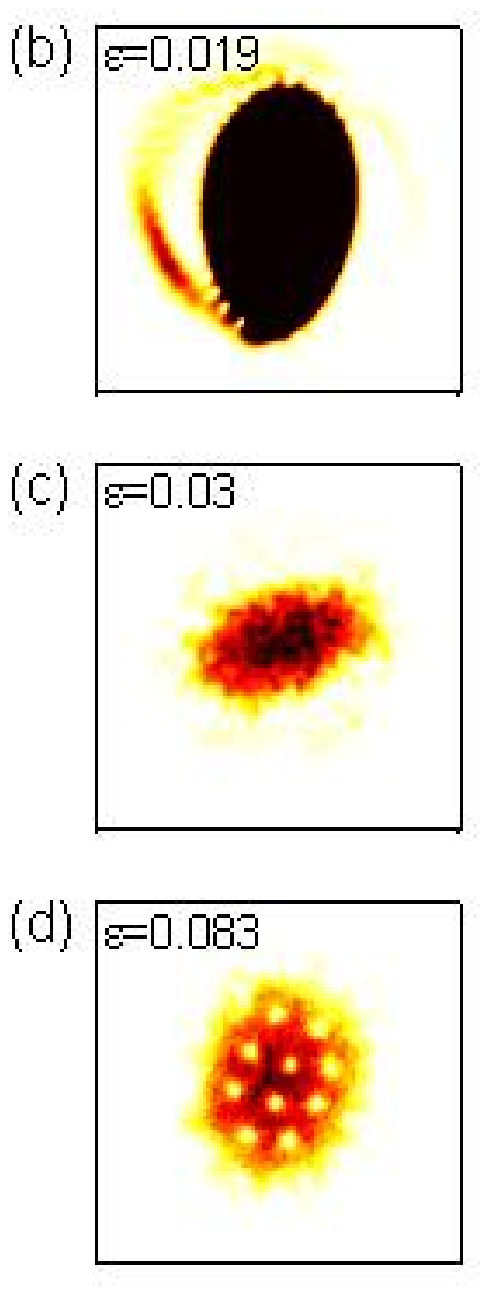}
\hspace{0.2cm}
\includegraphics[width=2.2cm,clip=true]{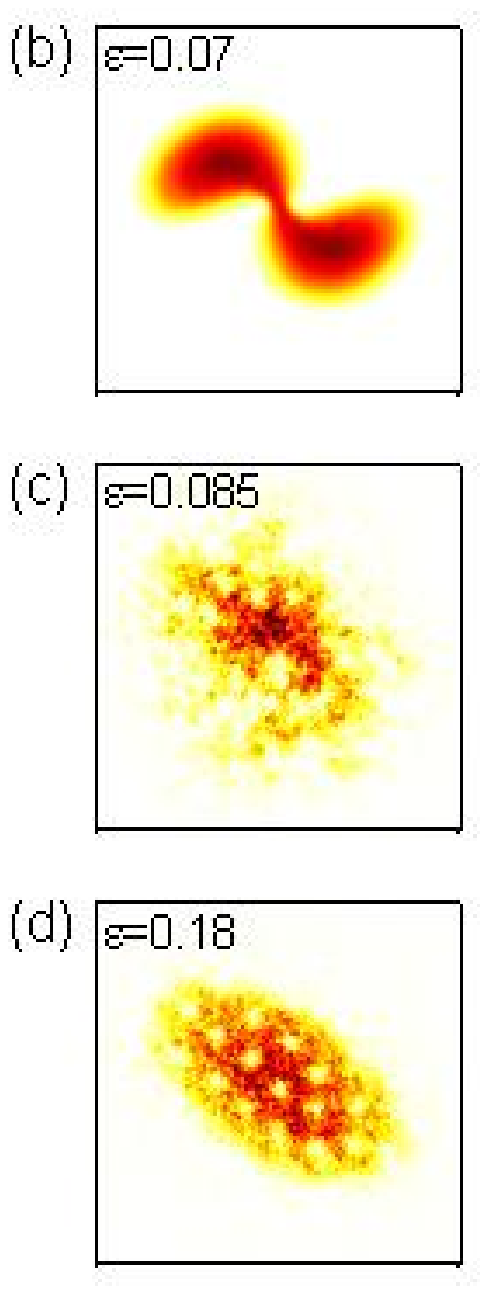}

\caption{Dynamics under a continuous increase of $\epsilon$ (at a
rate of $d\epsilon/dt=10^{-4}\omega_\perp$) for I. Ripple
instability ($\Omega/\omega_\perp=0.7$), II. Interbranch instability
($\Omega/\omega_\perp=0.75$), and III. Catastrophic instability
($\Omega/\omega_\perp=0.8$). (a) Velocity field amplitude $\alpha$
versus $\epsilon$ according to Eq.~(\ref{alphacontinuity}) (solid
lines) and TD simulations of Eqs.~(\ref{hydrodynamic1}) and
(\ref{hydrodynamic2}) (dots). In the latter case, $\alpha$ is
derived by fitting the velocity field to Eq.~(\ref{velocity}). To
the right of the dashed line this is no longer a valid fit (the
standard deviation of the fit becomes of the order of $\alpha$). The
regions of dynamical instability are indicated (circles). Density
snapshots corresponding to (b) the onset of instability, (c)
disrupted state seeded by the instability, and (d) a vortex lattice.
Once the instability point is reached, the dynamics are
qualitatively similar whether $\epsilon$ is continuously increased
or becomes fixed. Dark/light regions represent high/low density.
Each box represents a region $(12\times12)\mu m$ (the simulation box
is much greater than this). In I(b) and II(b) the density scale is
limited to $0.1\%n_0$ to highlight low density features. }
\label{fig2}
\end{figure}

{\em I Ripple instability}, $\Omega<\omega_\perp/\sqrt{2}$. The case
for $\Omega=0.7\omega_\perp$ is shown in Fig.~\ref{fig2}I(a). The
velocity field amplitude $\alpha$ (dots) follows the upper branch of
the rotating solutions (solid line), for which the condensate axes
rotate in phase with the trap axes, as noted experimentally
\cite{dalibard2,hodby}. The upper branch is always a solution, and,
as $\epsilon$ is increased, the condensate moves along the branch to
higher $\alpha$. However, when the ellipticity exceeds a critical
value $\epsilon_{\rm cr}$ (corresponding to when $\Omega>\Omega_{\rm
inst}$) the solution becomes dynamically unstable, according to
Eq.~(\ref{pertoperator}). For the example in Fig.~\ref{fig2}I(a),
$\epsilon_{\rm cr}\approx 0.09$ (dashed line). Subsequently $\alpha$
(dots) deviates from the rotating solutions of
Eq.~(\ref{alphacontinuity}) (solid lines), consistent with the onset
of dynamical instability. For low $\epsilon$, $\alpha$ (dots)
features small oscillations due to the centre-of-mass motion caused
by the initial offset of the condensate.

At a critical ellipticity $\epsilon_{\rm cr}^{\rm dev}$ low density
ripples form on the condensate edge (Fig.~\ref{fig2}I(b)), each on
the scale of the healing length and featuring a phase singularity
(`ghost' vortices \cite{tsubota,penckwitt,lobo}). These ripples grow
in amplitude as $\epsilon$ is increased. If $\epsilon$ becomes fixed
when the ripples have very low amplitude they do not grow over the
timescales considered. However, once $\epsilon$ exceeds a second
critical value $\epsilon_{\rm cr}^{\rm ins}$ (corresponding to when
the ripples have amplitude of order of $10\%n_0$), the dynamical
instability of Eq.~(\ref{pertoperator}) is triggered by the ripples.
This instability generates largescale shape oscillations
(Fig.~\ref{fig2}I(c)), disrupting the condensate, and enabling
`ghost' vortices to nucleate into the condensate, which slowly
crystallise into a lattice (Fig.~\ref{fig2}I(d)) \cite{Nick1}. Once
$\epsilon_{\rm cr}^{\rm ins}$ is reached, lattice formation occurs
independently of whether $\epsilon$ becomes fixed or continuously
increased.  Since the ripples that trigger the dynamical instability
originate in the non-TF tails of the BEC, they cannot be explained
by the TF analytics of Eqs.~(3) - (8), and a higher-order (non-TF)
approach would be required to explain their origin.

Surface ripples have been observed experimentally to precede vortex
nucleation at this rotation frequency \cite{hodby}. The gradual
growth of the ripples leads to a {\em slow} injection of
energy/angular momentum into the system, as has been observed in
previous studies within this frequency regime \cite{lobo,Emil}. Note
that according to Eq.~(\ref{pertoperator}) the dynamical instability
on the upper branch couples to odd modes only. If the symmetry is
preserved we do not expect the instability to develop
\cite{penckwitt,tsubota,lobo}.

{\em II Interbranch instability},
$\omega_\perp/\sqrt{2}<\Omega<\Omega_{\rm X}$. The case for
$\Omega=0.75\omega_\perp$ is shown in Fig.~\ref{fig2}II(a). Since
$\Omega>\omega_\perp/\sqrt{2}$, $\alpha$ (dots) initially follows
the lower branch solutions, where we observe that the BEC and trap
axes are $\pi/2$ out of phase. As $\epsilon$ is increased a point is
reached when $\Omega<\Omega_{\rm bif}(\epsilon)$. Here the lower
branch is no longer a solution of Eq.~(\ref{alphacontinuity}). For
the example shown, this occurs for $\epsilon \approx 0.02$ (dashed
line in Fig.~\ref{fig2}II(a)). Due to the non-TF nature of the
numerical solutions of Eqs. (\ref{hydrodynamic1}) and
(\ref{hydrodynamic2}), $\alpha$ does not perfectly fit the branch
solutions of Eq.~(\ref{alphacontinuity}).

When $\alpha$ (dots) reaches the cusp of the lower branch it
deviates non-adiabatically, as observed experimentally
\cite{dalibard2}. Since $\Omega<\Omega_{\rm X}$, the upper branch is
dynamically stable at this ellipticity. The condensate tries to
transform to the upper branch, but without dissipation it cannot
relax to this state. Instead $\alpha$ (dots) oscillates between
positive and negative values, and the condensate density undergoes
quadrupolar shape oscillations between an almost circular and highly
elliptical shape.  If $\epsilon$ becomes fixed close to this
critical ellipticity, the quadrupolar shape oscillations are stable.
However, if $\epsilon$ is increased further, the shape oscillations
destabilise, with the condensate shedding low density material at
its extrema in a spiral pattern (Fig.~\ref{fig2}II(b)). This
situation is closely analogous to when the rotation/ellipticity is
suddenly turned on, with the fate of the condensate being
qualitatively similar \cite{Nick1,Nick2}. The growth of the ejected
material gradually destabilises the condensate
(Fig.~\ref{fig2}II(c)), leading to vortex nucleation and ultimately
a lattice (Fig.~\ref{fig2}II(c)). This is fully consistent with the
observations in \cite{dalibard2}.

{\em III Catastrophic instability}, $\Omega>\Omega_{\rm X}$. The
case for $\Omega=0.8\omega_\perp$ is shown in Fig.~\ref{fig2}III(a).
Again $\alpha$ (dotted line) follows the lower branch, which ceases
to be a solution at some critical $\epsilon$ (dashed line). However,
since $\Omega>\Omega_{\rm X}$, the upper branch is dynamically
unstable at this point, and {\em no} stable solutions exist. The BEC
undergoes a quick and catastrophic instability, with $\alpha$ (dots)
deviating rapidly from the rotating solutions of
Eq.~(\ref{alphacontinuity}). The BEC density profile becomes
strongly contorted into a spiral shape (Fig.~\ref{fig2}III(b)). The
arms of the spiral collapse inwards and trap phase singularities to
form vortices. Energy and angular momentum are very rapidly injected
into the BEC (in contrast to the gradual ripple and interbranch
instabilities), as observed in \cite{lobo,Emil} for this frequency
regime.  The BEC becomes highly excited and turbulent
(Fig.~\ref{fig2}III(c)), with structure on length-scales less than
the healing length. Although we observe this state to ultimately
settle into a lattice (Fig.~\ref{fig2}II(c)), one may question the
validity of our zero-temperature condensate simulations for such a
`heated' state. Indeed, for $\Omega\gtsimeq 0.78\omega_\perp$,
turbulent states, rather than vortex lattices, were observed
experimentally in \cite{dalibard1}.

In the TD simulations we have measured two distinct critical
ellipticites: $\epsilon_{\rm cr}^{\rm dev}$ (crosses in
Fig.~\ref{fig1}(b)) is when, for a continuously increasing
$\epsilon$, we observe the density  to deviate from a smooth ellipse
(on the level of $0.1\%n_0$); $\epsilon_{\rm cr}^{\rm inst}$ (points
with error bars in Fig.~\ref{fig1}(b)) is when, for $\epsilon$
ramped up to some final value, instability and lattice formation
ultimately occur. In regime I, typically $\epsilon_{\rm cr}^{\rm
dev}\leq \epsilon_{\rm cr}^{\rm inst}$ since surface ripples are
generated which are stable for a narrow range of $\epsilon$, above
which they induce instability and lattice formation.  In regimes II
and III, $\epsilon_{\rm cr}^{\rm dev}\approx \epsilon_{\rm cr}^{\rm
inst}$, indicating the relatively sudden onset of this instability
once the density deviates from a smooth ellipse.

According to the TD simulations, the region above the points in
Fig.~\ref{fig1}(b) is unstable, leading to lattice formation. The
prediction of Eq.~(\ref{pertoperator}) (dashed line) gives
reasonable agreement with the TD results in region I, while
Eq.~(\ref{alphacontinuity}) gives excellent agreement in region III.
Also plotted in Fig.~\ref{fig1}(b) are the experimental results of
Hodby {\it et al.} \cite{hodby} (circles).  The TD results give good
agreement with the experimental data throughout, with the agreement
being particularly good in region III.

So far we have considered the adiabatic introduction of ellipticity
for fixed $\Omega$.  However, the results in Fig.~\ref{fig1}(b) also
apply to the case when $\Omega$ is introduced adiabatically for
fixed $\epsilon$.  Here the condensate always follows the upper
branch and so is only prone to the ripple instability when the upper
branch solution becomes dynamically unstable.  According to
Eq.~(\ref{pertoperator}) dynamical instability occurs when
$\Omega>\Omega_{\rm inst}(\epsilon)$ (dashed line in
Fig.~\ref{fig1}(b)). In \cite{dalibard2}, for $\epsilon=0.025$,
vortex lattice formation was observed to occur when $\Omega\gtsimeq
0.75\omega_\perp$, which agrees well with our TD simulations and
Fig.~\ref{fig1}(b).

Equations ~(\ref{alphacontinuity}) and (\ref{pertoperator}) are
valid in the TF limit of $\mu/\hbar\omega_\perp>>1$. For the TD
simulations of Eqs.~(\ref{hydrodynamic1}) and (\ref{hydrodynamic2}),
which do not make this assumption, we have employed an intermediate
value of $\mu/\hbar\omega_\perp=10$, but have tested other values.
For less (more) Thomas-Fermi-{\em like} condensates, the TD results
in Fig.~\ref{fig1}(b) shift downwards (upwards), deviating away from
(towards) the TF predictions of Eqs.~(\ref{alphacontinuity}) and
(\ref{pertoperator}).  In particular, as $\mu/\hbar\omega_\perp$ is
increased,  the TD values of $\alpha$ in Fig.~\ref{fig2} II-III(a)
(dots) become closer to the branch solutions (solid lines).

We have shown that vortex lattice formation is inherently a
two-dimensional and zero temperature effect. We have theoretically
mapped out the condensate stability as a function of rotation
frequency $\Omega$ and trap ellipticity $\epsilon$, with our
interpretation being consistent with previous experimental and
theoretical results. Specifically, for fixed $\Omega$ and adiabatic
introduction of $\epsilon$ we find three distinct regimes of
instability - {\em ripple} ($\Omega<\omega_\perp/\sqrt{2}$), {\em
interbranch} ($\omega_\perp/\sqrt{2}<\Omega<\Omega_{\rm X}$) and
{\em catastrophic} ($\Omega>\Omega_{\rm X}$). Each instability
manifests itself in a characteristic manner, which could be observed
experimentally. Ultimately in each case the instability seeds vortex
lattice formation.  This formation process not only relies on the
presence of an instability but is crucially dependent on the
breaking of the two-fold rotational symmetry of the system, as
inevitably occurs experimentally.

We acknowledge support from the ARC and the University of Melbourne.
We thank Y. Castin, C. S. Adams and C. J. Foot for helpful comments,
and C. J. Foot for the use of experimental data.


\begin{thebibliography}{99}
\bibitem{numerical_noise} Over long times, symmetry can break
intrinsically through the stimulated growth of numerical noise
\cite{Nick1,Nick2}.
\bibitem{narrow} Numerical simulations do not indicate any dynamical
instability within these narrow, low eigenvalue regions.
\bibitem{3d} As in \cite{Sinha} we find that the regions of
instability according to Eq.~(\ref{pertoperator}) are independent of
$\omega_z$.
\end{thebibliography}
\end{document}